\documentclass[aps,prl,groupedaddresss,superscriptaddress, twocolumn]{revtex4}
\usepackage{graphicx}
\usepackage{subfigure}
\usepackage{dcolumn}
\usepackage{bm}
\usepackage{amsfonts}
\usepackage{mathrsfs}
\usepackage{amssymb}
\usepackage{amsmath}
\usepackage{color}
\usepackage{chngcntr}
\usepackage{stmaryrd}
\usepackage[colorlinks=true,breaklinks=true,linkcolor=red,anchorcolor=red,citecolor=red,urlcolor=red]{hyperref}
\usepackage{booktabs}
\usepackage{multirow}

\begin{document}

\title{Transport measurement of fractional charges in topological models}

\author {Shu-guang Cheng}
\affiliation{Department of Physics, Northwest University, Xi'an 710069, China}

\author {Yijia Wu}
\affiliation{International Center for Quantum Materials, School of Physics, Peking University, Beijing 100871, China}

\author {Hua Jiang}
\email{jianghuaphy@suda.edu.cn}
\affiliation{School of Physical Science and Technology, Soochow University, Suzhou 215006, China}
\affiliation{Institute for Advanced Study, Soochow University, Suzhou, 215006, China}

\author {Qing-Feng Sun}
\affiliation{International Center for Quantum Materials, School of Physics, Peking University, Beijing 100871, China}
\affiliation{CAS Center for Excellence in Topological Quantum Computation, University of Chinese Academy of Sciences, Beijing 100190, China}

\author {X. C. Xie}
\affiliation{International Center for Quantum Materials, School of Physics, Peking University, Beijing 100871, China}
\affiliation{CAS Center for Excellence in Topological Quantum Computation, University of Chinese Academy of Sciences, Beijing 100190, China}

\begin{abstract}
The static topological fractional charge (TFC) in condensed matter systems is related to the band topology and thus has potential applications in topological quantum computation. However, the experimental measurement of these TFCs in electronic systems is quite challenging. We propose an electronic transport measurement  scheme that both the charge amount and the spatial distribution of the TFC can be extracted from the differential conductance through a quantum dot coupled to the topological system being measured. For one-dimensional Su-Schrieffer-Heeger (SSH) model, both the $e/2$ charge of the TFC and its distribution can be verified. We also show that the Anderson disorder effect, which breaks certain symmetry related to the TFC, is significant in higher-dimensional systems while has little effect on the one-dimensional SSH chain. Nonetheless, our measurement scheme can still work well for specific higher-order topological insulator materials, for instance, the $2e/3$ TFC in the breathing kagome model could be confirmed even in the presence of disorder effect.
\end{abstract}


\maketitle

{\it Introduction.--} The fractional charge, widely existed in a variety of topological systems, directly relates to the non-trivial topology of electronic states and is of great significance in the field of condensed matter physics \cite{Cui,Laughlin1,LaughlinRMP,FQHE1,Solitons,SSH2,SSH1,Wilczek,Schrieffer}. There are mainly two categories of fractional charge: the moving type and the static type. The former kind acts as charge carrier in the fractional quantum Hall (FQH) state \cite{Cui,Laughlin1,LaughlinRMP,FQHE1}. And the later one, can be found as topological edge states in many topological materials, such as one-dimensional systems with the Jackiw-Rebbi mechanism \cite{Solitons,SSH2,SSH1,Wilczek,Schrieffer},  topological crystalline insulators with disclination \cite{crystalline1}, and higher-order topological insulators \cite{Quadrupole1,Quadrupole2,tjModel,Kagome1,Kagome2,Zeng,Yao1}. Remarkably, both the moving \cite{FQHEEx1,non-Abelian1,non-Abelian2,non-Abelian3} and static fractional charges \cite{Yijia1,Yijia2,corner,QuantumCom1} follow the quantum statistics beyond the Boson/Fermion statistics and can be employed for topological quantum computation.

Back in 1997, the moving fractional charge in the FQH state was experimentally verified via the transport measurement of shot noise \cite{FQHE2,FQHE3}. The detection of the static topological fractional charge (TFC), in contrast, is quite challenging in condensed matter systems. The direct measurement of TFC via scanning tunneling spectroscopy (STS) is hindered by the experimental resolution \cite{crystalline2-3}. In addition to that, earlier theories have proposed an approach to detect the Jackiw-Rebbi zero mode through its $\pi$-period Aharanov-Bohm oscillation \cite{Yijia1,Trans_T1}. Although the presence of the zero mode could be certificated, the charge amount of the static TFC cannot be determined in such a proposal. In another earlier theoretical work \cite{QSHE}, a proposal is raised based on an effective model where the TFC trapped by the magnetic domains can be detected via the Coulomb blockade. Such a simplified scheme, in which the entire electronic structure as well as the disorder effect are excluded, has not been confirmed in a more realistic lattice model. Another drawback of this scheme is that the spatial distribution of the TFC cannot be obtained. It is worth noting that although the experimental measurement of TFC has recently been reported in classical wave systems \cite{crystalline2-3,crystalline4,crystalline2,crystalline3}, the elusive disorder effect remains to be further investigated. Significantly, the TFC state here is occupied only when an input with certain frequency is provided. The absence of Fermi surface hinders the verification of the quantum statistics of the TFC in classical wave systems.

In recent years, remarkable progresses for the experimental realization of Su-Schrieffer-Heeger (SSH) model in condensed matter systems have been made by engineering graphene nanoribbons \cite{Grphene_SSH12,Grphene_SSH1,Grphene_SSH2}. By precisely decorating the graphene nanoribbon edge profiles, both the topologically trivial and non-trivial states are manifested through the STS. Owing to these progress, the enthusiasm for the discrimination and measurement of TFC in topological materials has been highly raised.


In this theoretic work, the static TFC in topological systems is obtained through the electronic transport of a quantum dot (QD) coupled to the topological system. For the SSH model, both the $e/2$ TFC and its spatial distribution is obtained from transport results. The disorder effect, which is widely presented in topological systems supporting TFC but has not been thoroughly studied yet, is also intensively investigated in this work.
The Anderson disorder breaking certain symmetry related to the TFC generally has significant effect on the TFC. However, such effect is greatly suppressed in one-dimensional system that for disordered SSH chain, the TFC amount measured is still in good agreement with the clean result. For certain higher-order topological materials, e.g. some armchair-edged breathing kagome material, the well-localized TFC possessing $2e/3$ charge could also be confirmed by our transport scheme even under disorder.
\begin{figure}
\includegraphics[width=0.5\textwidth]{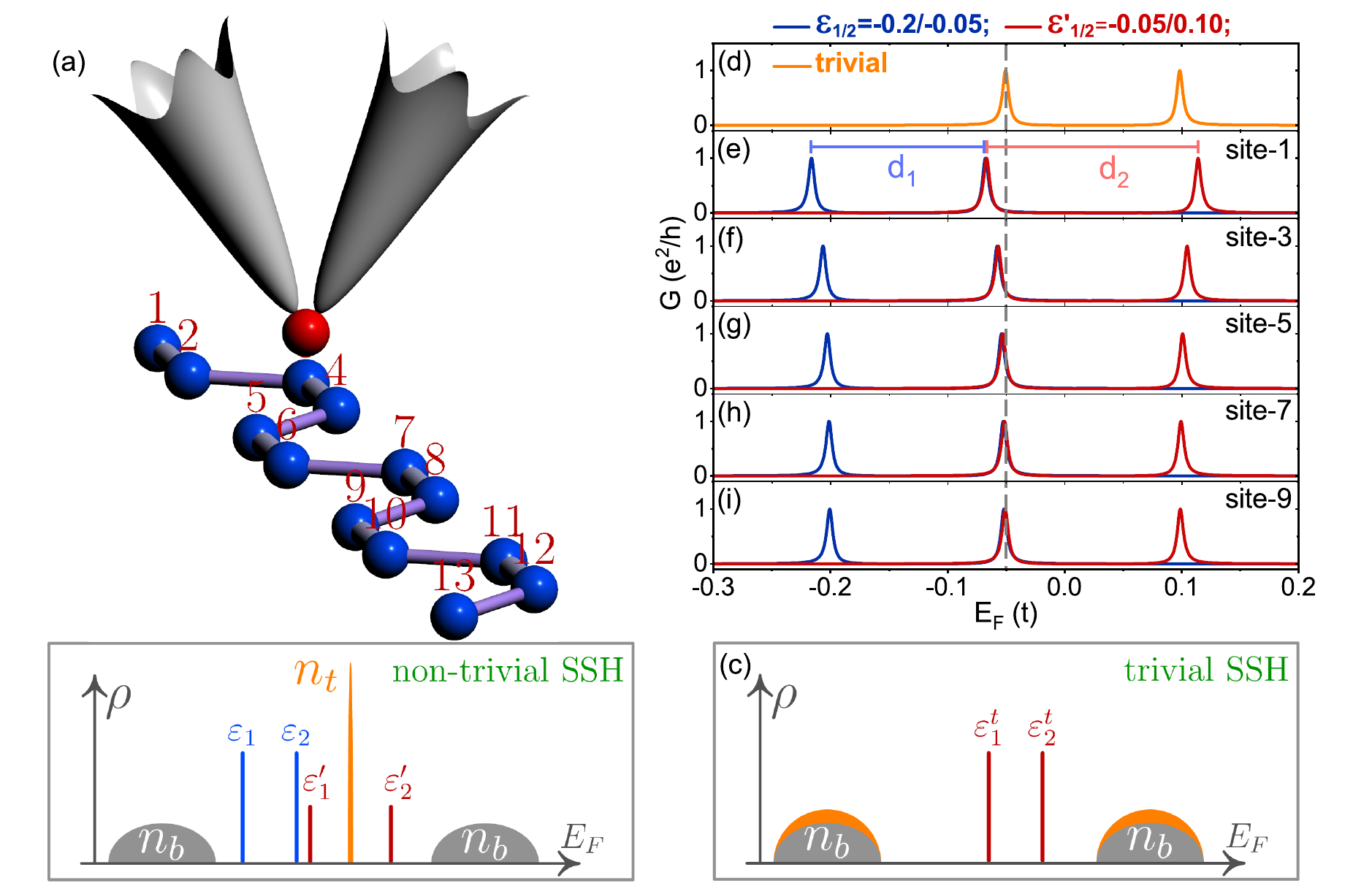}
\caption{
(a) Schematic plot of the transport device measuring the TFC in the SSH chain.
(b), (c) Band structure for non-trivial/trivial SSH model and the QD levels. The number of electron of the zero mode $n_t$ in (b) is the same as the increased portion of the bulk states' number of electron in (c), as denoted by the orange areas.
(d)-(i) Fermi level dependence of the conductance curves for (d) trivial end of the SSH chain with $t_1=0.6t$, and $t_2=0.37t$; and (e)-(i) non-trivial end of the same SSH chain with $t_1=0.37t$, and $t_2=0.6t$. Other parameters are $U=0.05t$, $t_k=0.1t$ and $t_c=0.01t$.
}\label{Fig1}
\end{figure}

{\it Model and methods.--} Our transport system
measuring TFC is shown as Fig.  1(a) that a QD (the red ball) with two energy levels is connected to two separated terminals (the grey ones). The QD can be moved to the vicinity of a specific atom of the topological system (e.g. a SSH chain shown by blue balls) and weakly bonds to this target atom, leading to a Coulomb interaction\cite{QD1,QD3,QD4,QD5}. In this way, the charge possessed by this specific target atom can be extracted from the shift of the differential conductance resonance peak\cite{QD2,QD6,QD7}. Then the amount and spatial distribution of the TFC possessed by the topological edge states can be extracted by measuring all the atoms involved.

There are two basic rules for this TFC measurement scheme: i) Any topological edge state carries an integer number of electrons. ii) A trivial insulator is charge neutral when the Fermi energy lies inside the gap. The first rule enables us to determine the Coulomb interaction strength, while the second rule enables us to count the charge by comparing the charge difference between a trivial insulator and its non-trivial counterpart. Remarkably, in the presence of disorder, whether  the second rule is still valid depends on the symmetry of the disorder term and the dimension of the topological material.


We first apply our method to the SSH chain composing of an odd number of atoms, in which the TFC is presented only in one end of the chain (non-trivial end), while another end of the chain (trivial end) behaves like a trivial insulator. In this way, the measurement and the charge difference comparison can be conducted in a single device. The total Hamiltonian of the system is in the form of:
\begin{eqnarray}\label{EQ1}
H=H_{\mathrm{QD}}+H_{\mathrm{SSH}}+H_{\mathrm{c}},
\end{eqnarray}
\noindent where the two-level QD (with level index $i$) is connected to the source and drain (with index $\alpha=L,R$) as $H_{\mathrm{QD}}=\sum_{k\alpha}\varepsilon_{k\alpha}C^\dagger_{k\alpha}C_{k\alpha}+\sum_{i}\varepsilon_{i}d^\dagger_{i}d_{i}+\sum_{k\alpha,i}t_{k}(C^\dagger_{k\alpha}d_{i}+h.c.) $.
$H_{\mathrm{SSH}}=\sum_{i}(t_1b_{2i-1}^\dagger b_{2i}+t_2b_{2i+1}^\dagger b_{2i}+h.c.)$ describes the SSH chain and $t_{1/2}$ is the alternative coupling between nearest atoms.
The coupling term reads $H_{\mathrm{c}}=\sum_i[Ud^\dagger_{i}d_{i}b^\dagger_{s}b_{s}+t_c(d_i^\dagger b_s+b_s^\dagger d_i)]$ with $U$ the Coulomb interaction strength, and $t_c$ the direct tunneling between the QD and the target atom (denoted by index $s$) of the SSH chain.
The linear conductance of the system is obtained from the Green's function \cite{Green1,Green2}
 \begin{eqnarray}\label{G}
\mathcal{G}(E) =\frac{e^2}{h}\mathrm{Tr}[\Gamma^r_LG^r\Gamma^r_R(G^r)^\dagger].
\end{eqnarray}
Here $\Gamma^r_{L/R}$ is the symmetric line-width function of the terminals, $G^r(E_F) $ is the retarded Green's function which needs to be solved by iteration for non-zero $U$ \cite{iteration}. When the QD is weakly coupled to the topological system, the resonance peaks of the $\mathcal{G}(E_F)$ locate at $\varepsilon_i+U\langle n_s\rangle$ where $\langle n_s\rangle$ is the total electron number below the Fermi level \cite{peak,Haug}. In practice, $U$ is not a prior known parameter which should also be extracted from the measurment.

{\it SSH model.--} For a QD consisting of two energy levels $\epsilon_1, \epsilon_2$, when both these two levels are below the fractionally-charged subgap zero mode of the SSH chain as $\varepsilon_1 < \varepsilon_2 < 0$ [Fig. 1(b)], the corresponding conductance peaks are separated by $d_1=\varepsilon_2-\varepsilon_1$. Alternatively, by tuning the gate voltage $V_g$, these two levels can be elevated to $\varepsilon_1'$ and $\varepsilon_2'$ that $\varepsilon_1'<0$ and $\varepsilon_2'>0$ [Fig. 1(b)]. Since $\varepsilon_1'=\varepsilon_1+eV_g$ and $\varepsilon_2'=\varepsilon_2+eV_g+Un_t$, where $n_t$ denotes the number of electron of the zero mode at the site being measured, now the conductance peaks are separated as $d_2=\varepsilon_2'-\varepsilon_1'=d_1+Un_t$.
Owing to the integer charge rule mentioned earlier, $\sum n_t=1$ when all the atoms that the TFC resides in are considered. Fig. 1 (e)-(i) display the $\mathcal{G}-E_F$ relations for all the odd sites at the non-trivial end of the SSH chain for both i) $\varepsilon_1, \varepsilon_2<0$; and ii) $\varepsilon_1'<0$, $\varepsilon_2'>0$. The conductance peak shifts for even atoms are trivial as $d_1=d_2$ (not shown)\cite{Supp}. Figure 2(a) summarizes the site $n$ dependence of $\Delta d_n=d_2-d_1$ extracted from Fig. 1(e)-(i). For odd sites, $\Delta d_n$ decreases as the target atom moves from the end to the bulk, and for all the even sites, $\Delta d_n=0$. Summing them up gives $U_0=\sum d_n=0.052t$, which is quite close to the input value of $U=0.05t$ and demonstrates that the prior unknown parameter $U_0$ can be obtained from the transport measurement.

\begin{figure}
\includegraphics[width=0.48\textwidth]{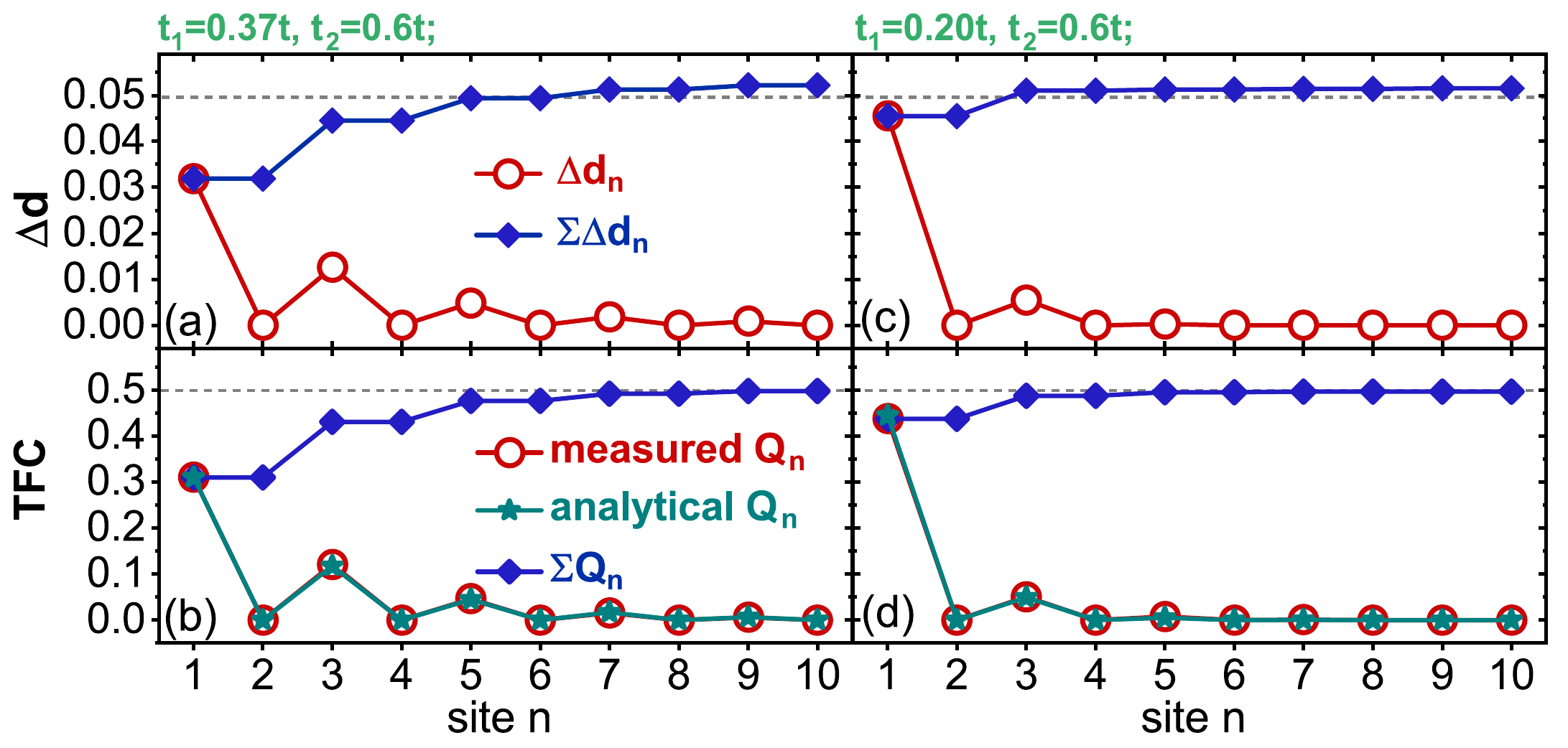}
\caption{
The TFC and its distribution in the SSH chain. (a), (c) The shift of the conductance peaks' separation $\Delta d_n=d_2-d_1$ [drawn from Fig. \ref{Fig1} (e)-(i)] and its summation $\sum_n\Delta d_n$. (b), (d) The TFC distribution $Q_n$ and its summation $\sum_n Q_n$.
}\label{Fig2}
\end{figure}

To figure out the amount and the distribution of the TFC, we need to measure the trivial end of the same SSH chain while turning off the gate voltage as $V_g=0$. Compared with the non-trivial end, the bulk states' electron number in the trivial end is increased by $n_t$ [see Fig. 1(c)]. The conductance peaks now locate at $\varepsilon_1^t$ and $\varepsilon_2^t$, where $\varepsilon_{1}^t = \varepsilon_{1} - U_0 Q_n/e$ and $Q_n$ is the TFC at site $n$. Therefore, $Q_n$ can be extracted from the conductance peak shift as $Q_n = (\varepsilon_{1}-\varepsilon_{1}^t) e/U_0$ since $U_0$ has been obtained previously.
Fig. \ref{Fig2}(c) displays the spatial distribution of $Q_n$ and its summation $\sum_n Q_n$. For odd sites, $Q_n$ decreases as the target atom moves from the end to the bulk of the SSH chain and for even sites, $Q_n$ is nearly zero. The $Q_n$ measured is in good agreement with the analytical result: $Q_n = (e/2)(t_1/t_2)^{n-1}[1-(t_1/t_2)^2]$ for the odd site, and $Q_n=0$ for the even site \cite{SSH_ana,Supp,SSH_Rb}. Finally, $\sum_n Q_n$ approaches $e/2$ [Fig. \ref{Fig2}(b)], confirming the 1/2 charge quantization of the TFC in the clean SSH model.

The SSH parameters adopted above $t_1=0.37t$ and $t_2=0.6t$ ($t=1eV$ is the energy unit) are drawn from a practical experiment where the SSH chain is built from the graphene naoribbon \cite{Grphene_SSH2}. Figure \ref{Fig2} (c)-(d) displays the TFC measured for another set of parameters $t_1=0.2t$ and $t_2=0.6t$ (see Supplementary Material for details \cite{Supp}). In such a case, $U_0=0.052t$ [Fig. \ref{Fig2}(c)] can be drawn from the transport data. The measurement scheme also shows that the TFC now becomes more localized at the end of the SSH chain, while the TFC measured is still close to $e/2$ [Fig. \ref{Fig2}(d)]. All these are in perfect match with the analytical results. Such transport measurement scheme is also adopted for obtaining the spatial distribution and verifying the $e/2$ amount of the TFC \cite{Supp} carried by the topological corner state of the quadrupole insulator \cite{Quadrupole1,Quadrupole2}, where the latter is regarded as the two-dimensional analogy of the SSH chain.

\begin{figure}
\includegraphics[width=0.55\textwidth]{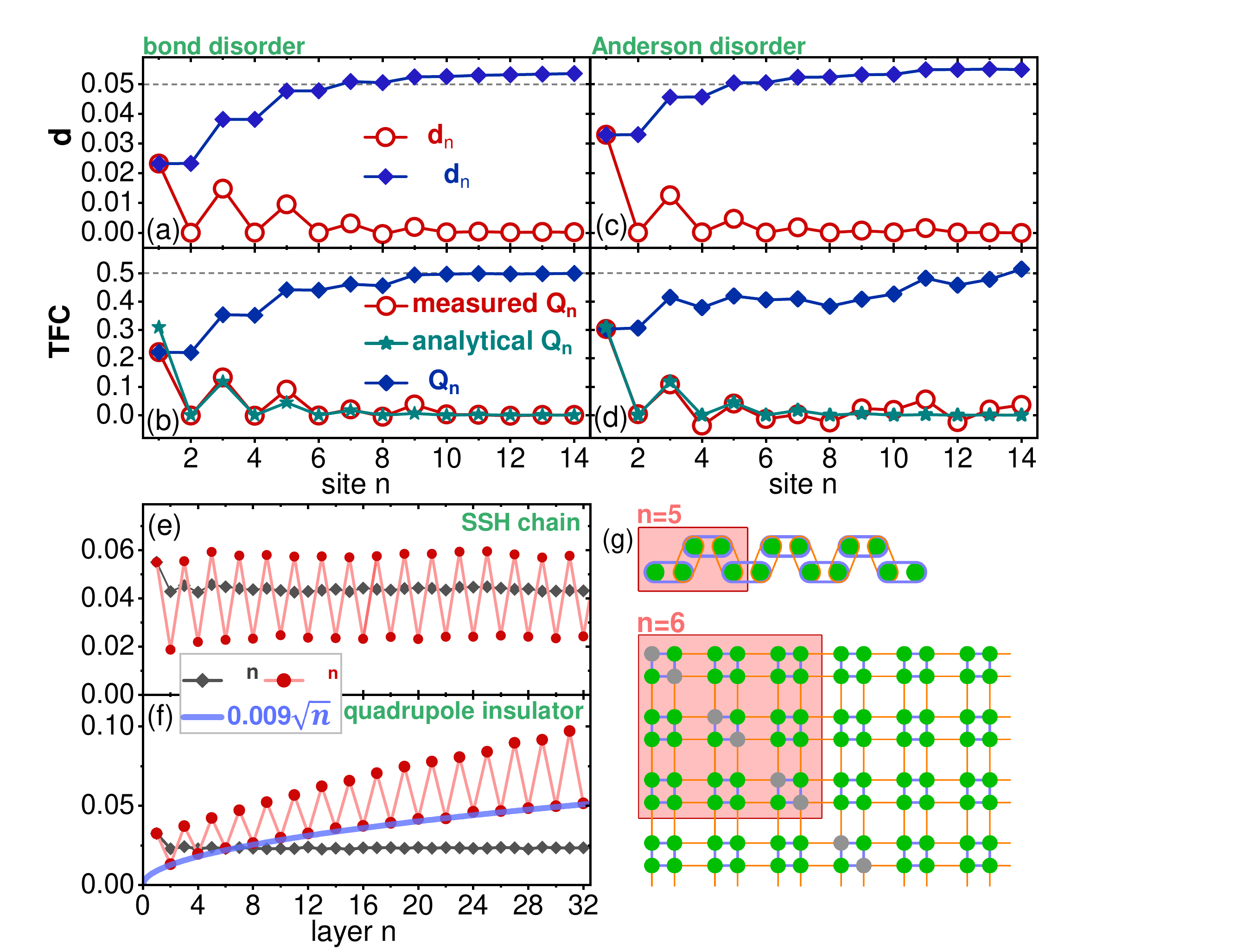}
\caption{
(a)-(d) The TFC distribution in the disordered SSH model as (a), (b) bond disorder with $W=0.2t$; (c), (d) Anderson disorder with $W=0.2t$. All other parameters are the same as those in Fig. \ref{Fig1}.
(e), (f) Standard deviation of the TFC $\sigma^n$ at site $n$, and standard deviation of the TFC inside the area concerned $\sigma^n_\Sigma$ under Anderson disorder for (e) SSH chain, and (f) quadrupole insulator. These two models are sketched in (g), where the red rectangles indicate the area being concerned. Both these two models have the same band gap ($\sim 0.4t$), sample length and disorder strength $W=0.5t$.
 }\label{Fig3}
\end{figure}

{\it Disorder effect.--} Disorder effect including bond disorder and Anderson disorder is widely presented in practical experiments, which may induce charge fluctuation and thus hinder the identification of the genuine fractional charge. The bond disorder, for instance, in the form of $\sum_{i} w_i(b_{i}^\dagger b_{i+1}+h.c.)$ in the SSH model where $w_i$ is uniformly distributed as $w_i\in [-W/2, W/2]$ and $W$ is the disorder strength \cite{bond_dis}, preserves the chiral symmetry obeyed by the clean SSH model. Hence the charge neutral rule for the trivial insulator remains valid, and the fluctuation only comes from the redistribution of the TFC in the non-trivial state. As shown in Fig. \ref{Fig3}(b), under bond disorder, although the distribution of the TFC deviates from the clean SSH model, the TFC amount still approaches $e/2$.

In contrast, the Anderson disorder $\sum_{i}w_ib_{i}^\dagger b_{i}$ \cite{And_dis} in the SSH model breaks the chiral symmetry so that the charge neutral rule is only satisfied on an average manner. The TFC fluctuation is now  ascribed to the redistribution of the electron density of  both the bulk states (in the trivial and non-trivial SSH chain) and the topological edge state (in the non-trivial SSH chain). The transport data confirms that the position of the conductance peaks in each site of the trivial SSH chain also fluctuates under Anderson disorder (see Fig. S3 in Ref. \onlinecite{Supp}), which is in stark contrast to the bond disorder condition. As a result, though the amount of the TFC is approximately $e/2$, both the TFC distribution $Q_n$ and its summation $\sum_n Q_n$ shows significant fluctuation [Fig. \ref{Fig3}(d)] even for the sites far away from the end of the SSH chain.

A question then arises that whether the fluctuation of the total TFC becomes more significant if more atoms are taken into consideration. This issue is essential because in higher-dimensional systems, the atom number involved for the topological edge state is proportional to the power of the localization length of the edge state.
For the one-dimensional SSH chain, the standard deviation of the TFC at each site $\sigma^n \equiv \sigma(Q_n)$ and the standard deviation of the total TFC inside the area concerned $\sigma_\Sigma^n \equiv \sigma(\sum_{i\in \square} Q_i)$ is investigated [``$\square$'' indicates the area concerned, shown by the red rectangles in Fig. \ref{Fig3}(g)].
Significantly, both $\sigma^n$ and $\sigma_\Sigma^n$ are in the same order.
As a comparison, as shown in Fig. \ref{Fig3}(f), (g), for an Anderson-disordered two-dimensional quadrupole insulator \cite{Quadrupole1,Quadrupole2} whose band gap and disorder strength are both the same as the one-dimensional SSH chain, although $\sigma^n$ is nearly independent of $n$ [$n$ refers the index of the grey dots in Fig. \ref{Fig3}(g)], the $\sigma_\Sigma^n$ here quickly increases in the fashion proportional to $\sqrt{n}$. Such fluctuation behavior is certainly detrimental to the TFC measurement.

In case of Anderson disorder, the fluctuations of $Q_n$ in adjacent sites are correlated so that the fluctuation of the total TFC inside the area concerned $\sigma_\Sigma^n$ is only determined by the charge fluctuation at the boundary of the area concerned.
For one-dimensional model like SSH chain, such a boundary is a single site, while for higher-dimensional systems like quadrupole insulator, the number of sites at the boundary 
increases with the area being concerned. Consequently, $\sigma_\Sigma^n$ does not increase with $n$ for one-dimensional topological system, while it quickly increases with $n$ for higher-dimensional system\cite{Supp}. In other words, for higher-dimensional topological materials supporting TFC, the disorder-induced fluctuation of the TFC is reduced for a better-localized topological edge state. For example, though the experiments have shown specific spatial crystal symmetries related to the TFC are broken by the inevitable disorder effect \cite{crystalline4,crystalline2,crystalline3}, the TFC could still be observed since the topological corner states here are well localized.



{\it Breathing kagome lattice.--} Recently, a TFC of $2e/3$ is reported in breathing kagome lattice \cite{Kagome1,Kagome2,Zeng,Yao1}, though such material is two-dimensional, the topological corner states therein can be well localized (e.g. monolayer $\mathrm{MoS_2}$\cite{Zeng,Yao1}), so that our transport measurement scheme is still applicable.
It is worth noting that in addition to the topological corner state, the zigzag-edged breathing kagome lattice \cite{Supp} also possesses a metallic one-dimensional edge state, hence the TFC here can not be detected by our scheme. Therefore, we first turn to investigate the armchair-edged breathing kagome lattice whose one-dimensional edge state is insulating\cite{Supp}.

\begin{figure}
\includegraphics[width=0.48\textwidth]{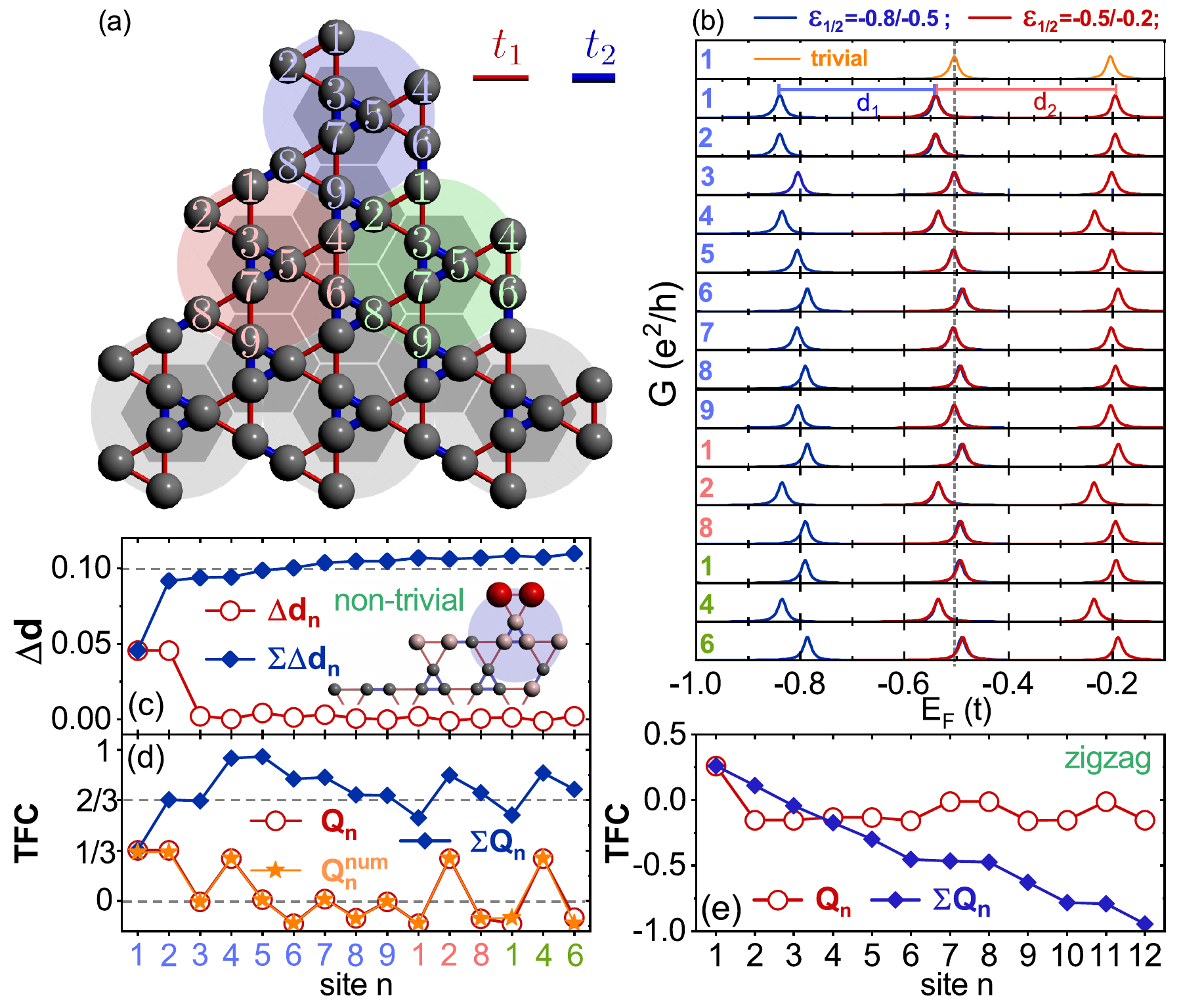}
\caption{
(a) Schematic plot of the armchair-edged breathing kagome lattice that the red (blue) bond indicates hopping amplitude $t_1$ ($t_2$). The atoms are numbered in each individual supercell as highlighted by colored circles. (b) The conductance curves in each site for both the trivial ($t_1=t, t_2=0.3t$) and the non-trivial ($t_1=0.3t, t_2=t$) cases.
(c) The conductance peak shift $\Delta d_n$ drawn from (b). Inset: the TFC is mainly distributed at the two sites marked by the red balls. (d) The TFC distribution $Q_n$ and $\sum_n Q_n$ drawn from (c), where the colored numbers in the horizontal axis refer to the site indices in (a). (e) The same TFC distribution for zigzag-edged breathing kagome model. Other parameters are $U=0.1t$, $t_k=0.1t$, and $t_c=0.01t$.
}\label{Fig4}
\end{figure}

Fig. \ref{Fig4}(a) shows a corner of a triangular armchair-edged breathing kagome lattice that the three supercells near the corner being mainly concerned are highlighted by colored circles.
Fig. \ref{Fig4}(b) exhibits the conductance curves obtained from our measurement scheme for the representative $15$ of the $27$ sites inside these three supercells. The conductance peak shifts $\Delta d_n$ are listed in Fig. \ref{Fig4}(c), and the TFC $Q_n$ as well as its summation are shown in Fig. \ref{Fig4}(d).
The  $Q_n$ obtained is in good agreement with the numerical result $Q^{num}_n$ by diagonalizing the Hamiltonian of an armchair-edged breathing kagome flake\cite{Supp,Eigen}.
It is shown that the TFC is mainly distributed in the edge sites [see inset of Fig. \ref{Fig4}(c)]. Moreover, at the edge, taking the site 1, 2, 8 in the pale red supercell as an example, although the TFC at each of these sites is non-zero, their summation makes no net contribution to the total TFC. The summation $\sum_n Q_n$ approaches $2e/3$, which confirms the quantized TFC as expected.

As a comparison [see Fig. \ref{Fig4}(e)], we also show that our transport measurement scheme fails to determine the TFC in the zigzag-edged breathing kagome lattice \cite{Kagome1}, because the ``charge neutral'' rule, one of the two basis rules of our measurement scheme, is broken by its metallic one-dimensional topological edge state\cite{Yao1,Supp}. Finally, the spin degeneracy has been ignored from the beginning. When the spin doubling is also taken into consideration, for instance, for the quadrupole insulator, one may be confused by the ``integer'' TFC of $2(e/2)$ \cite{crystalline2-3}. In contrast, the $2e/3$ TFC is always fractional even after considering the spin doubling, which serves as an additional advantage for the breathing kagome materials.

{\it Conclusion and Discussion.--} In summary, a transport measurement scheme is proposed to measure the amount and the spatial distribution of TFC in topological materials.
Through such a scheme, the $e/2$ amount of the TFC in the SSH model as well as its spatial distribution has been verified. 
In the presence of bond disorder preserving chiral symmetry, similar results can still be obtained.
It implies that seeking a material in which the symmetry related to the TFC is quite robust will facilitate the experimental identification of the quantized TFC.
In the presence of Anderson disorder breaking chiral symmetry, the fluctuation of the TFC amount is largely suppressed in one-dimensional systems. Meanwhile, for Anderson-disordered higher-dimensional topological materials such as breathing kagome lattice, the amount and the distribution of the TFC can still be managed to obtained for the well-localized topological corner states. It indicates that in specific condensed matter materials, the difficulty of experimentally distinguishing the genuine fractional charge and the disorder-induced fluctuation can be circumvented.

{\it Acknowledgement.--} We thank J. Zeng, H. W. Liu, and Z. Q. Zhang for fruitful discussions. This work is financially supported by National Basic Research Program of China (Grants No. 2019YFA0308403), the National Natural Science Foundation of China (Grant No. 11874298, No. 11822407),  the Strategic Priority Research Program of Chinese Academy
of Sciences (Grant No. XDB28000000), and China Postdoctoral Science Foundation (Grant No. 2021M690233).


\begin{widetext}
\newpage
\setcounter{equation}{0}
\setcounter{figure}{0}
\setcounter{table}{0}

\begin{center}
\textbf{Supplementary Materials for ``Transport measurement of fractional charges in topological models"}
\end{center}

\begin{center}
Shu-guang Cheng$^{1}$, Yijia Wu$^{2}$, Hua Jiang$^{3,4*}$, Qing-Feng Sun$^{2,5}$, and X. C. Xie$^{2,5}$
\end{center}

\begin{center}
$^1$~{\it Department of Physics, Northwest University, Xi'an 710069, China}

$^2$~{\it International Center for Quantum Materials, School of Physics, Peking University, Beijing 100871, China}

$^3$~{\it School of Physical Science and Technology, Soochow University, Suzhou 215006, China}

$^4$~{\it Institute for Advanced Study, Soochow University, Suzhou, 215006, China}

$^5$~{\it CAS Center for Excellence in Topological Quantum Computation, University of Chinese Academy of Sciences, Beijing 100190, China}
\end{center}


\section{Model and methods}

For the QD-SSH model displayed in S.\ref{model}, the Hamiltonian (1) in the main text is divided to two parts:
 \begin{eqnarray}\label{EQ1}
H_0=\sum_{i}\varepsilon_{i}d^\dagger_{i}d_{i}+\sum_i[Ud_i^\dagger d_ib_s^\dagger b_s+t_c(d_i^\dagger d_s+d_s^\dagger d_i)]
\end{eqnarray}
and
 \begin{eqnarray}\label{EQ2}
H_1&=&\sum_{k\alpha}\varepsilon_{k\alpha}C^\dagger_{k\alpha}C_{k\alpha}+\sum_{k\alpha,i}t_{k}(C^\dagger_{k\alpha}d_{i}+d^\dagger_{i}C_{k\alpha})+\sum_{i}t_1b_{2i-1}^\dagger b_{2i}+t_2b_{2i+1}^\dagger b_{2i}+h.c.
\end{eqnarray}

So in the figure S.\ref{model}, the area within the rectangle is described by $H_0$ and the rest is included by $H_1$ (including terminals, the rest part of SSH model and the coupling in between). The terminal we used is a lattice model which is favored in case of integral. Here $\varepsilon_{k\alpha}=-6t\cos(k)$ with $t$ the energy unit used in the whole work.

The current from the left terminal $L$ is obtained from the non-equilibrium Green's function technique,
 \begin{eqnarray}\label{curr}
J_L&=&\frac{ie}{\hbar}\langle[\sum_{k\alpha}C^\dagger_{k\alpha}C_{k\alpha}, H]\rangle=\frac{ie}{2h}\int(f_L-f_R)\mathbf{\Gamma^L}[\mathbf{G^r}(E)-\mathbf{G^a}(E)]dE.
\end{eqnarray}

\begin{figure}[h]
\includegraphics[width=0.6\textwidth]{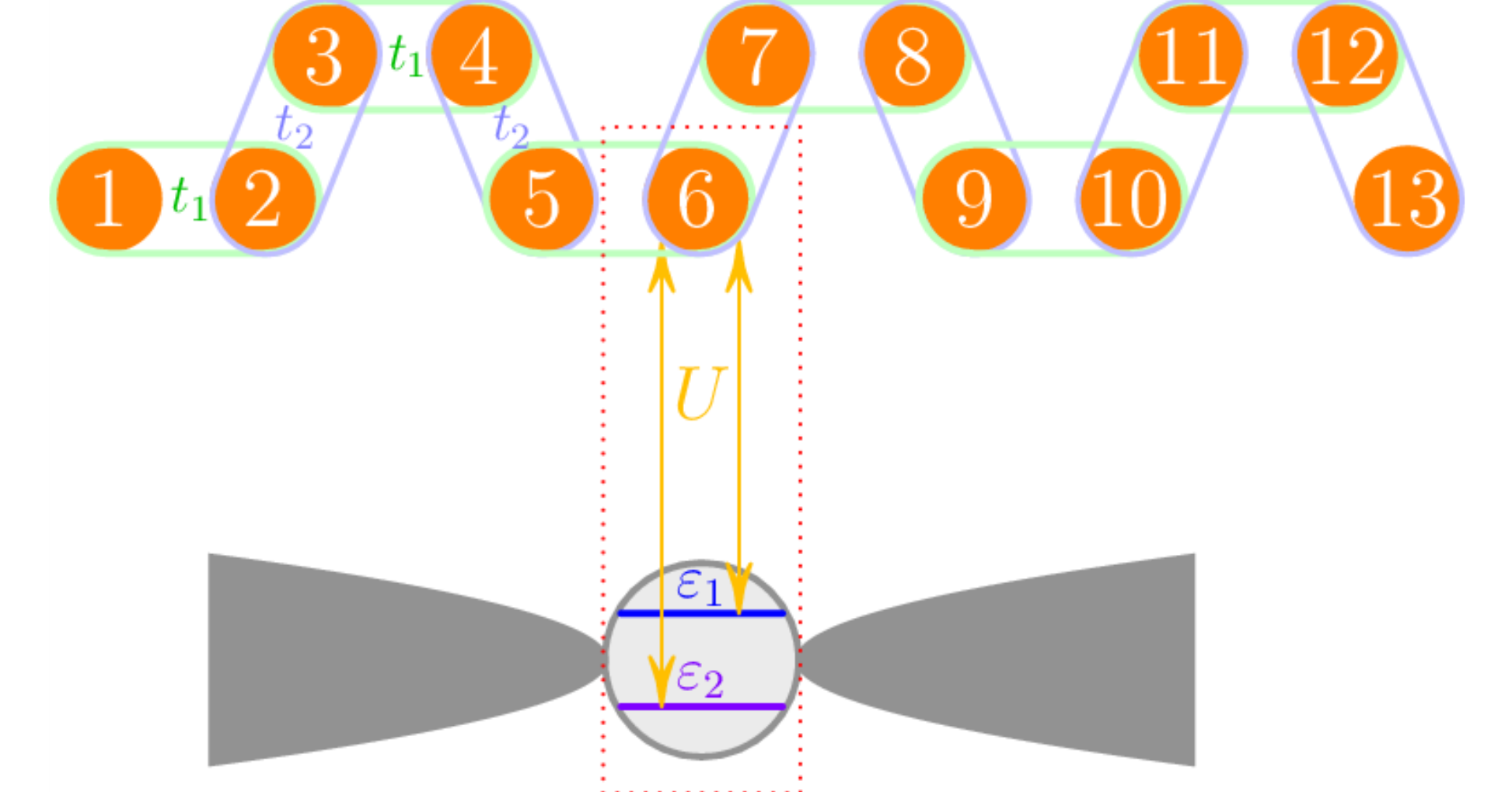}
\caption{
A schematic for QD-SSH model with a trivial end and a non-trivial end at separate ends. The coupling parameters of the SSH chain are $t_1$ and $t_2$. The QD, harbors two discrete levels $\varepsilon_1$ and $\varepsilon_2$, is bonded to the $6th$ atom in the SSH model via a weak coupling $t_c$ and Coulomb interaction strength $U$.
}\label{model}
\end{figure}

Here $f_\alpha=\frac{1}{1+e^{(E-E_F-eV_\alpha)/k_BT}}$ is the Fermi-Dirac distribution in terminal $\alpha$ and a low temperature is adopted.  $\mathbf{\Gamma^L}$ is the line-width function indicating the connection of the system (described by the equation \ref{EQ1}) to the environment. $\mathbf{G^r}(E)$ is the retarded green function calculated by

 \begin{eqnarray}\label{Gr}
\mathbf{G^r}(E)=
\left[
\left(
  \begin{array}{ccc}
    E-\varepsilon_{1}-Un_s & 0 & 0 \\
    0 &  E-\varepsilon_{2}-Un_s & 0 \\
    0 & 0 &  E-U(n_1+n_2) \\
  \end{array}
\right)-
\left(
  \begin{array}{ccc}
    \Sigma_1^r & \Sigma_1^r & t_c \\
    \Sigma_1^r &  \Sigma_1^r & t_c \\
    t_c & t_c &  \Sigma_3^r \\
  \end{array}
  \right)
    \right]^{-1}
\end{eqnarray}

In the above formula, the retarded self-energy is $\Sigma_1^r=t_k^2(g^r_L+g^r_R)$ with $g^r_\alpha(E)=\frac{1}{2\times(3t)^2}[E-\sqrt{E^2-4\times(3t)^2}]$ is the surface Green's function of terminal $\alpha$.
The treatment of self-energy $\Sigma_3^r$ in equation \ref{Gr} is a little more complicated. Taking figure S.\ref{model} as an example, the atom No. $6$ is connected to a segment of atom chain No.$1-5$ with coupling $t_1$ and a semi-infinite SSH model started from atom No. 7 with coupling $t_2$. So $\Sigma_3^r=t_1^2g^r_0(5,5)+t_2^2g^r_{SSH}$ with

 \begin{eqnarray}\label{G0}
g^r_0(E)=
\left(
  \begin{array}{ccccc}
    E & -t_1 & 0 & 0 & 0 \\
    -t_1 & E & -t_2 & 0 & 0 \\
    0 & -t_2 & E & -t_1 & 0 \\
    0 & 0 & -t_1 & E & -t_2 \\
    0 & 0 & 0 & -t_2 & E \\
  \end{array}
\right)^{-1}
\end{eqnarray}
and $g^r_{SSH}$ will be given in Eq. \ref{infini} in the next section. The particle numbers ($n_1$, $n_2$ and $n_s$) in equation \ref{Gr} refer to electron number in the quantum dot level $\varepsilon_1$, $\varepsilon_2$ and the atom they are connected to. They are obtained through solving the integral equation self-consistently by iteration, e.g. $n_s=-\frac{1}{\pi}\int_{-\infty}^{E_F} \mathrm{Im} [\mathbf{G^r}_{(3,3)}^r(E)]dE$ and the differential conductance is calculated from
 \begin{eqnarray}\label{G}
\mathcal{G}(E) =\frac{e^2}{h}\mathrm{Tr}[\mathbf{\Gamma^r_L}\mathbf{G^r}\mathbf{\Gamma^r_R}(\mathbf{G^r})^\dagger],
\end{eqnarray}
 in which $\mathbf{}=it_k^2[g^r_\alpha-(g^r_\alpha)^*]
 \left(
  \begin{array}{ccc}
    1 & 1 & 0 \\
    1 &  1 & 0 \\
    0 & 0 & 0 \\
  \end{array}
  \right)$.

\begin{figure}[b]
\includegraphics[width=0.6\textwidth]{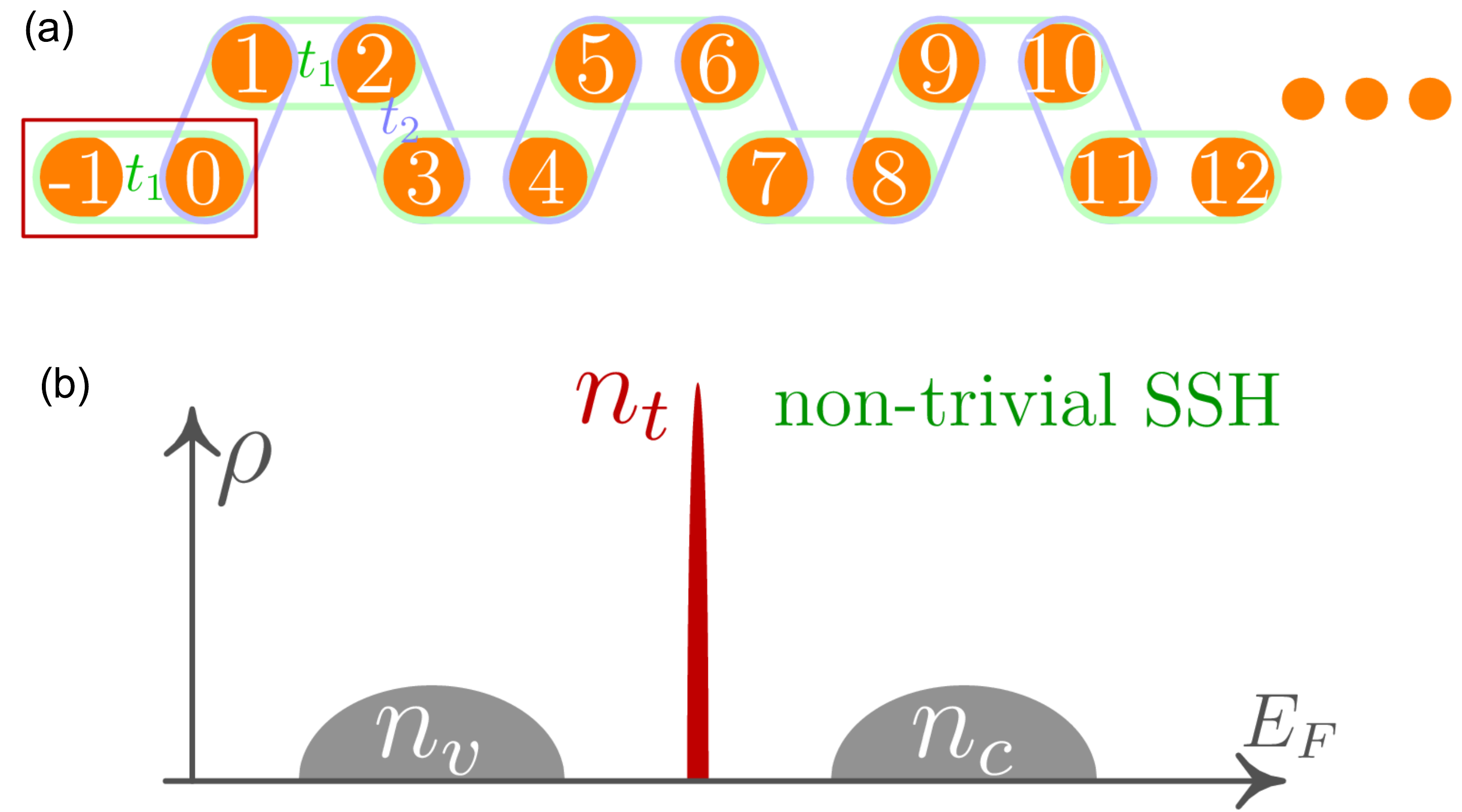}
\caption{
(a) The schematic for a semi-infinite SSH model. The atoms labeled No. 0 and No. -1 are augmented for the convenience of discussion. (b) The band structure for the non-trivial SSH chain, $n_v$, $n_t$ and $n_c$ are electron numbers for valence band, TFC and conduction band.
}\label{SSH_infinite}
\end{figure}

\section{Analytical Derivation of TFC in a SSH model}

\subsection{Clean model}
Now we solve the spatial distribution of TFC in a simple SSH model (see the model in S.\ref{SSH_infinite}) via Green's function method. For any given site $i$ in the SSH model, the electron number for valence band, conduction band and TFC (if non-zero) are $n_v$, $n_c$ and $n_t$, as displayed in S.\ref{SSH_infinite} (b). Guaranteed by the chiral symmetry, one has $n_v=n_c$ and $2n_v+n_t=1$. In case of a trivial (non-trivial) state, $n_t'=0$ ($n_t\neq0$) and $n_v'=1/2$ ($n_v=1/2-n_t/2$). Especially, under non-trivial case, when the Fermi level $E_F$ is within the gap and $E_F<0$, the charge of the site $i$ is $Q_i=-e(n_v-n_v')=en_t/2$. Alternatively, when the subgap Fermi level $E_F$ is above the zero energy, the charge is $Q_i=-n_te+en_t/2=-en_t/2$.

In the following, we derivate the site dependent of $Q_i$ using Green's function method\cite{Green}.
In S.\ref{SSH_infinite} (a), when the atoms in the red rectangle is absent, site 1 is the end of the semi-infinite SSH chain. We assume the retarded Green's function of the system as $G^r_{i,j}(E)$.
When the two extra atoms are added (No. 0 and -1), site No. -1 become the new end, and the Green's function $G^r_{-1,-1}(E)$ can be obtained via the Dyson equation,
\begin{eqnarray}\label{Dyson}
[E+i\eta-\frac{t_1^2}{E+i\eta-t_2^2G^r_{11}(E)}]G^r_{-1,-1}(E)=1
\end{eqnarray}
with $\eta$ a positive infinitesimal number.
Here $G^r_{-1,-1}(E)$ should be equals to $G^r_{11}(E)$. Hence we have
 \begin{eqnarray}
t_2^2[G^r_{1,1}(E)]^2-(E+i\eta-\frac{t_1^2-t_2^2}{E+i\eta})G^r_{1,1}(E)+1=0.
\end{eqnarray}
Solve the above equation one has
 \begin{eqnarray}\label{infini}
G^r_{1,1}(E)=\frac{1}{2t_2^2}[E+i\eta-\frac{t_1^2-t_2^2}{E+i\eta}-\sqrt{(E+i\eta)^2+(\frac{t_1^2-t_2^2}{E+i\eta})^2-2(t_1^2+t_2^2)}].
\end{eqnarray}
It gives $g^r_{SSH}$ in the above section. The density of states of the bulk electrons is given by
 \begin{eqnarray}
\rho(E)=-\frac{1}{\pi}\mathrm{Im}[G^r_{1,1}(E)]=\frac{1}{2\pi t_2^2}\sqrt{2(t_1^2+t_2^2)-E^2-\frac{(t_1^2-t_2^2)^2}{E^2}}.
\end{eqnarray}
One can easily notice that $\rho(E)=\rho(-E)$, thus $n_v=n_c$, in accordance with the chiral symmetry. In the following discussion, we use the model in S.\ref{SSH_infinite} (a) when the atoms in the rectangle is absent, i.e. the atoms of a semi-infinite SSH model are labeled from No.1.

Before calculating the TFC, we first collect the retarded self-energy functions near $E=0$ for the sake of convenience:
\begin{eqnarray}
\begin{cases}
\Sigma_1^r = t_1^2(E+i\eta)^{-1} \nonumber \\
\Sigma_2^r = t_2^2[E+i\eta-\Sigma_1^r]^{-1}=-t_2^2(E+i\eta)/t_1^2 \nonumber \\
\Sigma_3^r = t_1^2[E+i\eta-\Sigma_2^r]^{-1}=t_1^4/(t_1^2+t_2^2)/(E+i\eta)  \\
\Sigma_4^r = t_2^2[E+i\eta-\Sigma_3^r]^{-1}=-t_2^2(t_1^2+t_2^2)(E+i\eta)/t_1^4 \nonumber \\
\Sigma_5^r = t_1^2[E+i\eta-\Sigma_4^r]^{-1}=t_1^6/(t_1^4+t_2^4+t_1^2t_2^2)/(E+i\eta) \nonumber \\
\cdots
\end{cases}
\end{eqnarray}

\textbf{Site No.1}: The amount of the TFC in site 1 can be obtained from Eq. \ref{infini} by setting $E\rightarrow 0$,
 \begin{eqnarray}\label{zero}
G^r_{1,1}(E\rightarrow 0)=\frac{t_2^2-t_1^2+|t_2^2-t_1^2|}{2t_2^2(E+i\eta)}.
\end{eqnarray}

i) When $t_1>t_2$, $G^r_{1,1}(E\rightarrow 0)=0$ and the semi-infinite SSH is under trivial state;  ii) When $t_1<t_2$, $G^r_{1,1}(E\rightarrow 0)=(1-{t_1^2}/{t_2^2}){\frac{1}{E+i\eta}}$ and SSH is under non-trivial state. Meanwhile the density of state is $\rho(E)=(1-{t_1^2}/{t_2^2})\delta(E)$, and $n_t(1)=\int_{0^-}^{0^+}\rho(E)dE=1-{t_1^2}/{t_2^2}$ for the first site. Hereinafter we focus on the non-trivial case ($t_1<t_2$).

There are two kinds of SSH models (i.e. the trivial one and the non-trivial one), and they can switch to each other by removing the end site.
According to labels in S.\ref{SSH_infinite} (a), we name them ``odd" type and ``even" type. In the above discussion, $G^r_{1,1}(E)$ is actually ``odd" type: $G^r_{odd}=G^r_{1,1}(E)$.
The Green's function of site $1$ and site $2$ are associated by $G^r_{odd}(E)=[E+i\eta-t_1^2{G}^r_{even}(E)]^{-1}$.
So we have the retarded Green's function ${G}^r_{even}(E)$ for site No. $2$ if site 1 is removed: ${G}^r_{even}(E)=(E+i\eta)/(t_1^2-t_2^2)$. Collecting them together, we have
\begin{eqnarray}
\begin{cases}
G^r_{odd}=(1-{t_1^2}/{t_2^2})/{(E+i\eta)} \nonumber \\
G^r_{even}=(E+i\eta)/(t_1^2-t_2^2)  \nonumber
\end{cases}
\end{eqnarray}
for $E\rightarrow 0$.

\textbf{Site No.2}: For the second site, since $t_1<t_2$, we have
 \begin{eqnarray}\label{No2}
G^r_{2,2}(E\rightarrow 0)=[E+i\eta-\Sigma_1^r-t_2^2G^r_{odd}(E)]^{-1}=[E+i\eta-\frac{t_2^2}{E+i\eta}]^{-1}.
\end{eqnarray}
Thus $\rho(E\rightarrow0)=0$ and $n_t(2)=0$.

\textbf{Site No.3}: $G^r_{3,3}(E)$ is accessed by
 \begin{eqnarray}\label{No3}
G^r_{3,3}(E\rightarrow0)=[E+i\eta-\Sigma_2^r-t_1^2{G}^r_{even}]^{-1}=\frac{t_1^2}{t_2^2}(1-\frac{t_1^2}{t_2^2})\frac{1}{E+i\eta}.
\end{eqnarray}
Accordingly, $\rho(E\rightarrow0)=\frac{t_1^2}{t_2^2}(1-\frac{t_1^2}{t_2^2})\delta(E)$ and $n_t(3)=\frac{t_1^2}{t_2^2}(1-\frac{t_1^2}{t_2^2})$.

\textbf{Site No.4}: $G^r_{4,4}(E)$ is accessed by
 \begin{eqnarray}\label{No4}
G^r_{4,4}(\rightarrow0)=[E+i\eta-\Sigma_3^r-t_2^2{G}^r_{odd}]^{-1}=-(t_1^2+t_2^2)(E+i\eta)/t_2^4.
\end{eqnarray}
It gives $\rho(E\rightarrow0)=0$ and $n_t(4)=0$.

\textbf{Site No.5}: $G^r_{5,5}(E)$ is
 \begin{eqnarray}\label{No5}
G^r_{5,5}(\rightarrow0)=[E+i\eta-\Sigma_4^r-t_1^2{G}^r_{even}]^{-1}=t_1^4(t_2^2-t_1^2)/(E+i\eta)/t_2^6.
\end{eqnarray}
It gives $\rho(E\rightarrow0)=\frac{t_1^4}{t_2^4}(1-\frac{t_1^2}{t_2^2})\delta(E)$ and $n_t(5)=\frac{t_1^4}{t_2^4}(1-\frac{t_1^2}{t_2^2})$.

\textit{Summarizing the above results, $n_t(i)$ is zero for even sites, and $n_t(i)=(\frac{t_1}{t_2})^{i-1}(1-\frac{t_1^2}{t_2^2})$ for odd sites. The site dependent TFC is $Q_i=en_t(i)/2$ and their combination gives $\sum_i(Q_i)=e/2$ .}
\begin{figure}[b]
\includegraphics[width=1\textwidth]{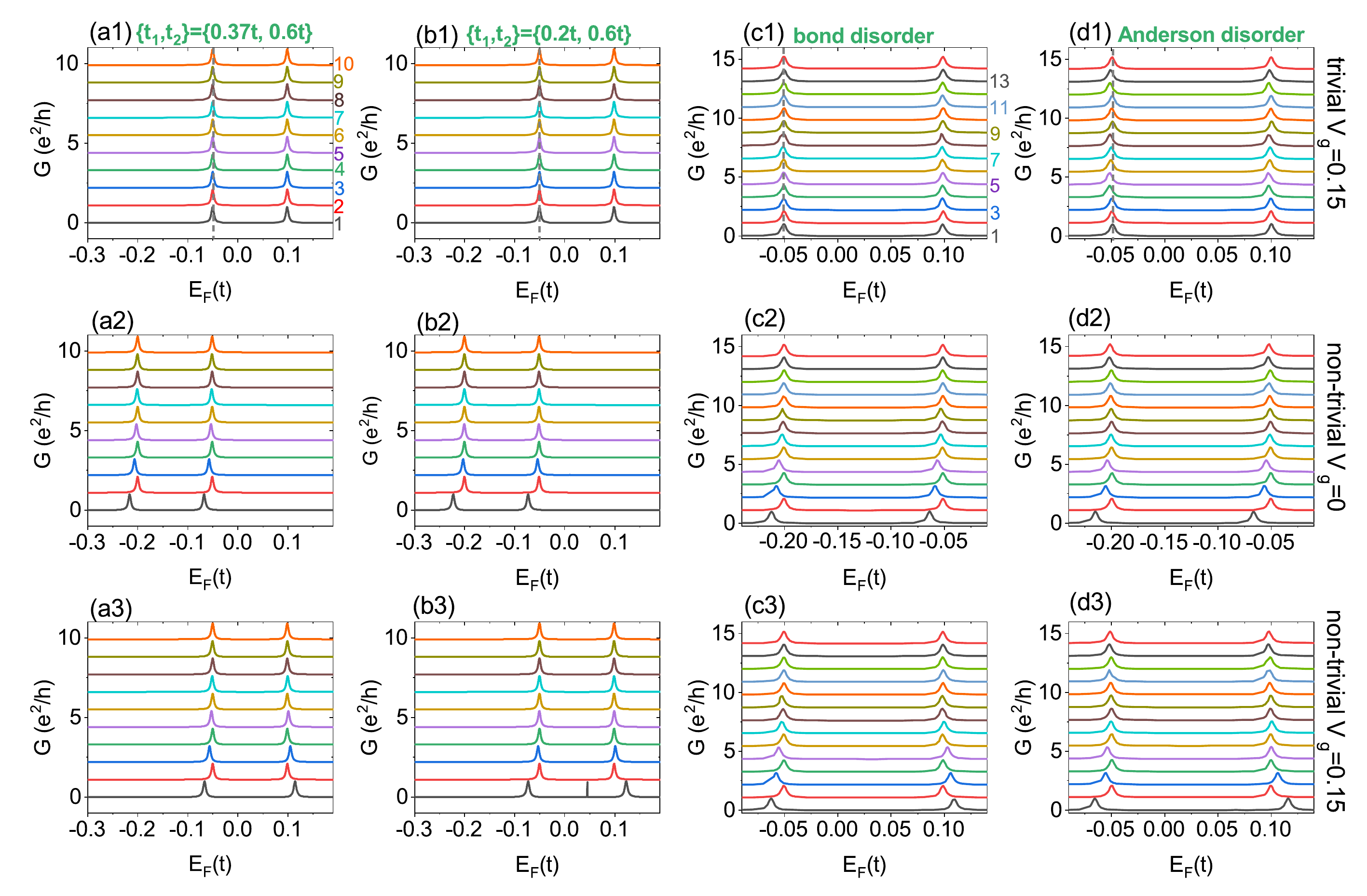}
\caption{
Transport data $\mathcal{G}$ vs $ E_F$ for the SSH-QD system in S.\ref{model}. (a1-a3): a clean SSH chain with parameters $\{t_1, t_2\}=\{0.37t, 0.6t\}$; (b1-b3) a clean SSH chain with parameters $\{t_1, t_2\}=\{0.2t, 0.6t\}$; (c1-c3) a SSH chain with bond disorder; (d1-d3) a SSH chain with Anderson disorder. For the disorder cases, $\{t_1, t_2\}=\{0.37t, 0.6t\}$ are sued. For each category, three cases are displayed: trivial state (the first column: a1, b1, c1, d1), non-trivial state with $V_g=0$ (the second column: a2, b2, c2, d2) and non-trivial state with $V_g=0.15t$ (the third column: a3, b3, c3, d3). In each figure, the curves for atoms from No.1 to No. 10 are arranged from bottom to top.
}\label{SSH}
\end{figure}

\subsection{Bond disorder and charge redistribution}
Now we discuss the effect of bond disorder on TFC. For a semi-infinite SSH chain under non-trivial state ($t_1<t_2$), if the first coupling (between site 1 and 2) $t_1$ is replace by $t_1+w$ with other couplings unchanged, we in the following calculate the TFC in site 1 and 2. By using the method in the previous subsection,
 \begin{eqnarray}\label{Dis_No1}
G^r_{1,1}(E\rightarrow 0)=[E+i\eta-(t_1+w)^2G^r_{even}(E)]^{-1}=\frac{t_2^2-t_1^2}{t_2^2+w^2+2wt_1}\frac{1}{E+i\eta}
\end{eqnarray}
We have the amount of TFC at site 1: $n_t(1)=\frac{t_2^2-t_1^2}{t_2^2+w^2+2wt_1}$. Clearly, the amount of TFC is affected by the bond disorder. Likewise,
 \begin{eqnarray}\label{Dis_No2}
G^r_{2,2}(E\rightarrow 0)=[E+i\eta-\frac{(t_1+w)^2}{E+i\eta}-t_2^2G^r_{odd}(E)]^{-1}=-\frac{E+i\eta}{t_2^2+w^2+2wt_1}
\end{eqnarray}
and $n_t(2)=0$. Thus under bond disorder, the TFC for the even site of SSH chain is still zero, as determined by the chiral symmetry. The result can also be observed from Fig.3 (b) in the main text.

\section{The transport data of SSH chain}
In S. \ref{SSH}, $\mathcal{G}$ vs $E_F$ curves for a SSH-QD system with parameters $\{t_1, t_2\}=\{0.37t, 0.6t\}$ and $\{t_1, t_2\}=\{0.2t, 0.6t\}$ are shown. The data analyzed from these data are displayed in the Fig. 2 of the main text. To calculate the TFC, the conductance peak shift for each site should be obtained from the conductance peaks of trivial state and nontrivial states. In the Fig.1 of main text, all the conductance peaks for non-trivial state are compared with a single curve for trivial state. It is guaranteed by the fact that in the trivial state, the electron states for all sites are the same and the position of the conductance peak stays unchanged [see S. \ref{SSH} (a1) and (b1)]. For non-trivial case, when both QD levels are less than the zero mode [displayed in S. \ref{SSH} (a2) and (b2)], both conductance peaks shift to a same direction and the value of the shift is proportional to the TFC $Q_n$. Alternatively, when the zero mode is located between two QD levels [S. \ref{SSH} (b3) and (d3)], two conductance peaks are shifted to opposite directions with a same value.

The transport data for SSH-QD system under disorders (the bond disorder and Anderson disorder) are shown in S. \ref{SSH} as well. Especially, under bond disorder, the conductance peaks for all sites in the trivial case stay the same due to the chiral symmetry [see the guideline in S.\ref{SSH} (c1)]. On the other hand, under Anderson disorder, in S.\ref{SSH} (d1), the conductance peaks fluctuate when the site varies. It means the charge neutral condition under Anderson disorder is only satisfied in an average manner. The redistribution of the bulk states, i.e. charge fluctuation, will lead to a measuring error. This effect exists broadly in all kinds of TFC systems, but is not well investigated.

\begin{figure}[b]
\includegraphics[width=0.8\textwidth]{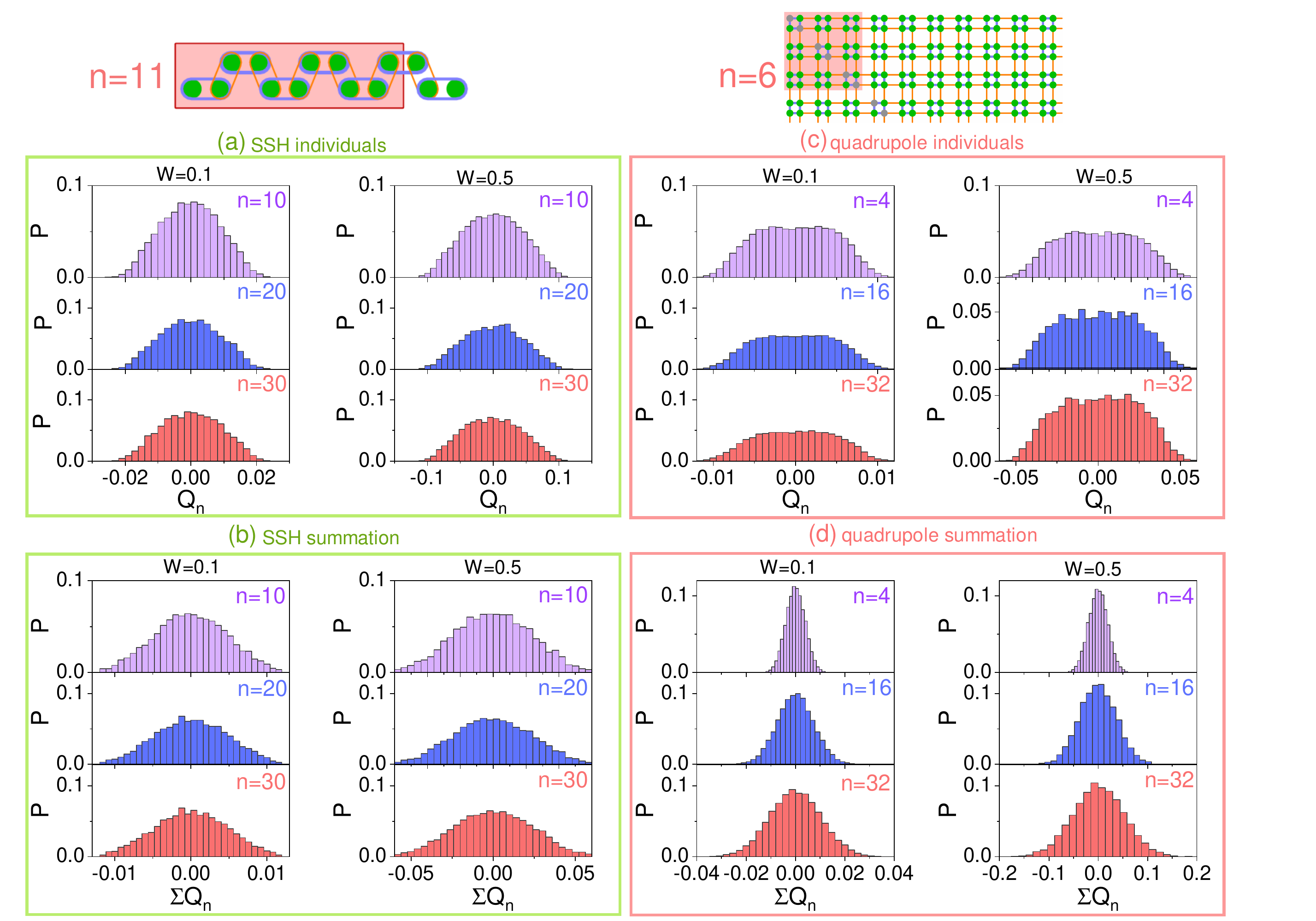}
\caption{
Net charge distribution of a SSH model (a-b) and a quadrupole insulator (c-d) for different disorder strength. Both the distribution for individual sites (a and c) and their summation (b and d) are displayed. For example, on the top of panel (a)/(c), the atoms within the light green/red shadow area are considered for summation, corresponding to $n=11$ and $n=6$, respectively. The sample parameters for both a SSH model and a quadrupole insulator are selected that they have a same energy gap ($\sim0.4t$): i) $t_1=1.0t$ and $t_2=0.8t$ for SSH model; ii)  $t_1=1.0t$ and $t_2=0.86t$. The parameters for the models are used that they have a same energy gap $\sim 0.4t$.
}\label{SSH_Hist}
\end{figure}
\section{sample dimension related net charge distribution}\label{disorder}
To investigate the sample dimension effect of charge fluctuation for topological materials under trivial state, the tight-binding model of the sample is studied.
We solve the eigen-equations $H\Psi_i=E_i\Psi_i$ in the presence of Anderson disorder. By summing up the states below the Fermi level of site $n$, the total charge $\mathcal{Q}_n$ is obtained by $\mathcal{Q}_n=e\sum_{i\in occ}|\Psi_i(n)|^2$\cite{Eigen}.
The net charge is obtained by $Q_n=Q_0-\mathcal{Q}_n$ with $Q_0$ the charge of nucleus.  In a clean trivial insulating system, $Q_n=0$ for all sites is guaranteed by the charge neutral condition. For SSH model or quadrupole insulator, $Q_0=e/2$. Here two cases are studied, the distribution of individual atoms and their summation. Specifically, in the top left panel of S.\ref{SSH_Hist}, the green area ($n=2$) in the SSH model, the distribution of $Q_n$ in the right most atom is considered for ``individual" case, and the summation of $Q_n$ within the whole green region is considered for the ``summation" case. Similarly, in the top right panel, the schematic of quadrupole insulator is displayed. In the red area, the distribution of $Q_n$ from the specific atom in the bottom-right corner is considered for ``individual" case. The choice of atom here is quite casual, in fact any atom is qualified and we do so is just for convenience. The summation of $Q_n$ within the whole red region is considered for the ``summation" case.

Figure S.\ref{SSH_Hist} shows the frequency histograms of net charge $Q_n$ distribution (a and c) and its summation $\sum Q_n$ (b and d) for both SSH chain (a-b) and quadrupole insulator (c-d) under different disorder strength. For SSH chain, the distribution of $Q_n$ subjects to a normal distribution and the standard deviation is independent of size No. $n$. Same features are found for stronger disorder only with a large standard deviation [see the data for $W=0.5t$ in S.\ref{SSH_Hist}(a)].

Now we check the ``summation" case in S.\ref{SSH_Hist}(b).
Intuitively, if a quantity comes from the summation of $n$ quantities with independent normal distributions (with the standard deviation $\sigma_0$), the summation possess a normal distribution with the standard deviation $\sigma=\sqrt{n}\sigma_0$.
In contrast, two interesting issues are found for the distribution of $\sum Q_n$. i) Although more atoms are involved (e.g. $n=30$), the standard deviation $\sigma^n$ stays the same. ii) Compared with the ``individual" case, the standard deviation $\sigma_\Sigma^n$ is even smaller. Both issues can be explained by the correlated electron redistribution in SSH model in which the disorder induced net charge redistribution $Q_n$ is not independent due to the strong bonding. Actually, although more atoms are involved in the statistics, only the outmost atom matters.
\begin{figure}[b]
\includegraphics[width=0.8\textwidth]{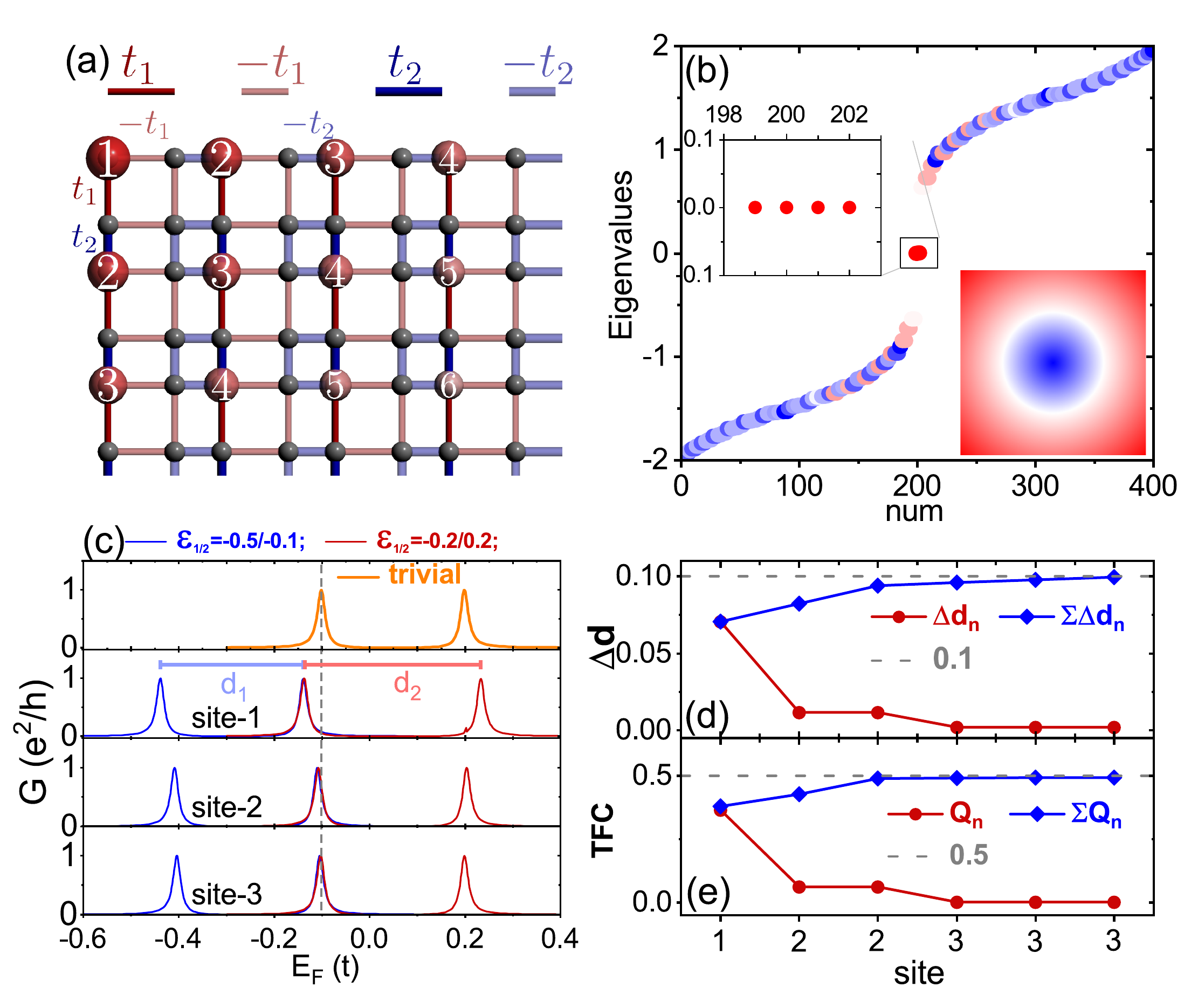}
\caption{
(a) The schematic for two-dimensional SSH model and the numbered red balls signify topological charge and the charge amount decreases from the corner. (b) The eigenvalue for a two-dimensional SSH model ($t_1=0.4, t_2=t$) with $20$ atoms in each side of a square nanoflake. The color of the dots signifies the average position of the eigenstate as shown in inset, thus the zero modes are corner states. (c) $\mathcal{G}$ vs $E_F$ relations for trivial ($t_1=t, t_2=0.4t$) and non-trivial ($t_1=0.4t, t_2=t$) cases in different sites. (d) Site dependence of $\Delta d_n$ and it summation adopted from (c). The dashed line is for $U=0.1t$. (e) Topological charge $Q_n$ and its summation $\sum_1^nQ_i$.
}\label{SSH_2D}
\end{figure}
Now we repeat the same process in quadrupole insulator in S.\ref{SSH_Hist} (c-d). The distribution of individual site share similar characteristics to those of SSH model. However, the distribution of $\sum Q_n$ shows a special feature: when more atoms are counted in the red rectangle, the standard deviation $\sigma^n_\Sigma$ is enhanced accordingly [see S.\ref{SSH_Hist} (d)]. It can be explained following the argument in the above discussed SSH model. For a two-dimensional model, when a larger sample is considered, the number of the outmost atoms is proportional to the sample size (e.g. $n=6$ in the up right panel of S.\ref{SSH_Hist}). Thus the effective atom number contributing the net charge fluctuation is proportional to the sample size $n$. Consequently, the standard deviation $\sigma_\Sigma^n$ is proportional to $\sqrt{n}$ as we have displayed in Fig. 3 (f) of the main text.

\section{TFC in quadrupole insulator}
The quadrupole insulator is the two-dimensional counterpart of SSH model. Its Hamiltonian in $k$ space reads\cite{Quadrupole1}:
 \begin{eqnarray}\label{quadrupole}
H(k)=[t_1+t_2\cos(k_x)]\tau_x\sigma_0-[t_1+t_2\cos(k_y)]\tau_y\sigma_y-t_2\tau_y\sigma_z \sin(k_x)-t_2\tau_y\sigma_x \sin(k_y),
\end{eqnarray}
where $\tau$ and $\sigma$ are Pauli matrices. Figure S.\ref{SSH_2D}(a) displays the scheme of a corner of the sample and the amount of TFC is denoted by the red ball.
The numbers attached to the red balls are used to mark the TFC. For a square shaped sample, the eigenvalues are shown in S.\ref{SSH_2D}(b). Here a square nanoflake with 20 atoms in each side is used, so there are totally $20\times20$ atoms/eigenvalues. The insets highlight the zero mode and the spatial distribution of the states is represented by color, the four degenerate zero mode come from four corner states. Both the trivial and non-trivial state of quadrupole insulator is measured with our scheme introduced in the main text and the results are shown in S.\ref{SSH_2D}(c). Since the performances of atoms with a same number are identical, here in the figure only three sets of curves are displayed (including six atoms). By collection them together, the results are shown in S.\ref{SSH_2D}(d). Similar to the SSH chain, the total TFC in quadrupole insulator is $e/2$ as well.
\section{TFC in zigzag edged breathing kagome model}
\begin{figure}[h]
\includegraphics[width=0.8\textwidth]{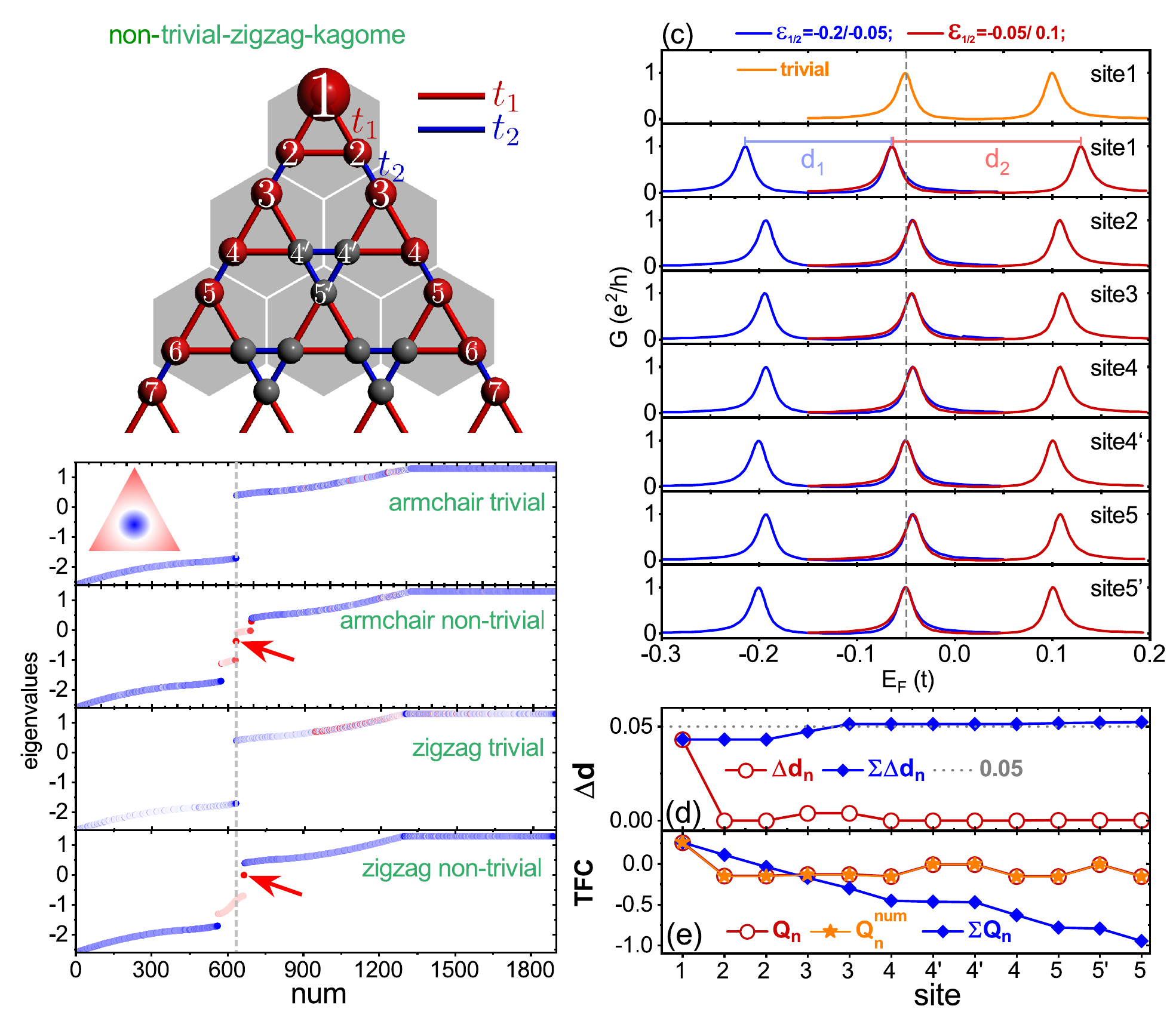}
\caption{
(a) Schematic of a tip of triangle-shaped zigzag edged breathing kagome nanoflake. The distribution and amount of TFC are indicated by the red balls. (b) The eigenvalues for an armchair edged and a zigzag edged breathing kagome model (both trivial and non-trivial states $\{t_1, t_2\}=\{0.3t, t\}$) with the same number of atoms. The grey dashed line locates at the $1/3$ position of the total atom number. The zero modes in both cases are highlighted by the red arrows. The color of the dots signifies the average position of the eigen-state as shown in inset. (c) $\mathcal{G}-E_F$ relations for trivial ($t_1=t, t_2=0.3t$) and non-trivial ($t_1=0.3t, t_2=t$) cases in different sites. (d) Site dependence of $\Delta d_n$ and $\sum\Delta d_n$ adopted from (c). The dashed line is for $U=0.05t$ (e) Topological charge $Q_n$ and $\sum_1^n Q_i$ for the model in (a), adopted from (c) and (d). $Q_n^{num}$ comes from the charge discrepancy between $1/3$ (the charge of nucleus) and the electronic states under occupation: $Q^{num}_n=e[1/3-\sum_{i\in occ}|\Psi_i(n)|^2]$, utilizing the same treatment as in section \ref{disorder}.
}\label{Kagome}
\end{figure}
Now we check the charge neutral condition for both armchair-edged and zigzag-edged breathing kagome models. In figure S.\ref{Kagome}(b), the eigen-energies of both models with a triangle-shaped nanoflake are studied under trivial and non-trivial states. The color of the dots means the position of the corresponding eigenstate [see the inset of S.\ref{Kagome}(b)]. The corner states in both cases are highlighted by red arrows. In case of an armchair-edged breathing kagome model under trivial state, there are $1/3$ eigenstates below the energy gap (see the grey dashed line). When the system becomes non-trivial, there are also nearly $1/3$ eigenstates below the corner mode. Since the Fermi level locates within the energy gap, the charge neutral condition guarantees the validity of our measurement scheme and hence we are able to obtain the $2e/3$ TFC in the main text.

The zigzag edged breathing kagome model [see S.\ref{Kagome}(a)] was a pioneer in the family of higher-order topological insulator\cite{Kagome}. Similar to a quadrupole insulator in the above discussion, in the figure S.\ref{Kagome}(a), the position and the amount of TFC is labeled by red balls and the results are obtained from the same diagonalization method. Alternatively, in S.\ref{Kagome}(b), although there are $1/3$ eigenstates below the gap under the trivial state, there are more than $1/3$ states below the corner states. The Fermi energy crossing the edge states, means the non-trivial breathing kagome model is actually metallic.
We apply our QD measurement to the zigzag edged breathing kagome model and the transport results are shown in S.\ref{Kagome}(c). Since the system in S.\ref{Kagome}(a) bears a mirror symmetry, the atoms under consideration is No. 1, 2, 3, 4, $4'$, 5 and $5'$ (their contribution is satisfactory and other atoms are not included). By collecting them together, in figure S.\ref{Kagome}(c), one notice TFC is mainly contributed by atom No. 1, 3, 5. Their summation gives $U_0\sim U$. So with our scheme the amount of Coulomb interaction $U$ is obtained successfully.

Finally, we study the distribution of TFCs. In S.\ref{Kagome}(e), the atoms focused is categorized to three kinds: i) site 1 and 3 carry the main part of the total TFC; ii) the charges carried by site 2, 4, 5 may be attributed to topological edge states. iii) site $4'$ and $5'$ are nearly of charge neutral and they carry no TFC.
The TFC $Q_n$ obtained here is in good agreement with the numerical result $Q_n^{num}$ by solving the eigen-equation of a zigzag-edged breathing kagome flake [see S.\ref{Kagome} (a)]\cite{Eigen}.
In S.\ref{Kagome}(e), the total TFC $\sum Q_n$ increases monotonously when more atoms (especially the ones at the edges) are included. The intervention of edge states hinders us from extracting the correct TFC from such samples, thus the present scheme fails.

\end{widetext}

\end{document}